# Generation and Measurement of High-order Optical Vortex via Cross-Phase

**CHEN WANG,[1] YUAN REN,[1,2,*] TONG LIU,[1] CHUANKAI LUO,[3] SONG QIU,[1] ZHIMENG LI,[1] AND HAO WU[1]**

[1]*Department of Aerospace Science and Technology, Space Engineering University, Beijing 101416, China*
[2]*State Key Lab of Laser Propulsion & its Application, Space Engineering University, Beijing 101416, China*
[3]*School of Physics and Optical Engineering, Xidian University, Xi'an, 710071, China*
*\*renyuan_823@aliyun.com*

**Abstract:** the generation and measurement of optical vortex (OV) are the basis for a variety of related applications. However, the special case of high-order OVs has not been sufficient addressed yet. Herein, a generation and measurement method of high-order OV via utilizing the CP is investigated. In the experiment, we generate OVs with $l = 60$, $p = 20$ and successfully measure OVs with $l = 200$, $p = 0$, where experimental results agree well with simulation outcome. On this basis, the intensity distributions of LG and HG beams (corresponding to the generation and measurement) versus waist radius of initial light beams is discussed. This work provides an alternative method to generate or measure high-order OV, which will facilitate applications in optical micro-manipulation, quantum entanglement and rotation speed detection.

## 1. Introduction

An Optical vortex (OV) with the spiral wavefront is a kind of structured light which carries an orbit angular momentum (OAM) per photon of $l\hbar$, where topological charge $l$ (TC) and radial node $p$ (RN) are two key parameters thereof. In 1989, Coullet et al. proposed a kind of optical field bearing some analogy with the superfluid[1]. In 1992, Allen showed that photons can carry OAM[2], which has aroused the interest of researchers. Since then, OV has been utilized in plethora of applications in the field of optical micro-manipulation[3], quantum entanglement[4] and rotation speed detection via the optical rotation doppler effect[5, 6], where increasing the TC or RN of OV is an important factor to improve the performance of applications. For instance, increasing TCs and RNs hold the potential to improve the transmission capacity in quantum information processing, and large TCs are also helpful to enhance the accuracy and sensitivity in the field of optical micro-measurement. Nonetheless, rare work considers situations of OVs with large TCs and RNs even though that would be of great importance for practical applications, and it's a real challenge to generate and measure OV with large TCs and RNs, i.e., *high-order optical vortex*[7-9].

    At present, the commonly used generation ways of OV are based on the spiral phase plate (SPP), the spatial light modulator (SLM) or fibers[10]. As far as these three are concerned, the fabrication of a SPP corresponding to high-order OV demands higher machining accuracy, the generation of high-order OV requires a SLM of high resolution and fibers can't be employed to generate high-order OV for now.

    Many methods have been proposed to measure the TCs of OV and relevant measurement methods can be mainly sorted into three categories[11, 12]: One type is interference-based which can be achieved by optical interferometry between OV and a plane wave, a spherical wave or a self-mirror wave[13], Other is diffraction-based which can be done via illuminating

OV on an aperture with specific shapes or a grating with the given phase structure[14]; another is based on the mode conversion[15], which has good robustness, can be realized with two cylindrical lens. However, the first two methods as mentioned above are not suitable for the measurement of OV with large TCs due to the dense distribution of interference fringes or diffraction lattices and the third method requires high accuracy of experimental setup which is complicated.

Especially, similar to the measurement of the large TCs, there are no ideal method to measure the large RNs even if the method based on the mode conversion could handle this but it's still hampered by the drawbacks which are same as the measurement of OV with large TCs as described earlier.

Recently, the cross phase (CP), a kind of phase structure, has been involved in LG(Laguerre-Gauss) beams and HG(Hermite-Gaussian) beams that opens up a new horizon for generation and measurement of high-order OV[11].

LG beams are a set of solutions to the paraxial wave equation in cylindrical coordinates while HG beams are another set of solutions in Cartesian coordinates. By adding the CP, the interconversion between two beams above can be done after propagating at a certain distance, which means these two kinds of beams are converted to non-eigenmodes, performing in radially dependent rotation during the propagation. This interconversion can be engaged in generation and measurement of LG beams. The generation and measurement via CP are not confined to LG beam, but also suitable for arbitrary high-order OV. For simplicity and without loss of generality, typically, we only consider the situation that the generation and measurement of high-order OV happen in the interconversion between LG beams and HG beams.

The method with the CP and the method of the mode conversion are both based on the astigmatism principle, while the former avoids the use of optical elements such as cylindrical lenses but holograms instead, which is more conducive to the precise manipulation of the light field and greatly evades the harsh requirements of relative position for cylindrical lenses.

In this paper, we investigate a generation and measurement method of high-order OV through utilizing the CP, which has been experimental achieved and the experimental results agree well with simulation outcomes. The distributions of LG and HG beams (corresponding to the generation and measurement) versus waist radius of initial light beams is discussed. To our best knowledge, it has not been reported on the experimental generation and measurement of high-order OV via the CP.

## 2. Theory

The form of the CP $\psi$ in Cartesian coordinates $(x, y)$ is

$$\psi = u(x\cos\theta - y\sin\theta)(x\sin\theta + y\cos\theta) \quad (1)$$

where the coefficient $u$ controls the conversion rate, and the higher the value of $u$, the faster the conversion rate. The azimuth factor $\theta$ characterizes the rotation angle of converted beams in one certain plane. It is noteworthy that Eqs.(1) could be simplified to $\psi' = uxy$ when $\theta = 0$ and we only take this typical situation into count in this article. On one hand, we can regard that the light field carrying the CP is a new type of light field. On the other hand, we can say that the CP is attached to the light field at a certain plane in the process of propagation. we would like to choose the latter view[11].

### A. Generation of high-order OV

We firstly consider the generation of high-order OV in the situation where the HG beams carry the CP. When $z_0 = 0$, the light field can be expressed as

$$U_G(x_0, y_0, 0) = H_n(x_0) H_m(y_0) \exp\left(-\frac{x_0^2 + y_0^2}{\omega_0^2}\right) \times \exp(i\psi) \quad (2)$$

where $(x_0, y_0, 0)$ denotes the initial plane, $H(\ )$ denotes the Hermite Gaussian polynomial, $n$ and $m$ are mode numbers, $\omega_0$ is the waist radius.

According to the Fresnel diffraction integral, when the light field mentioned above propagates a certain distance $z$, the output can be expressed that

$$E(x, y, z) = \frac{1}{i\lambda z} \exp(ikz) \exp\left(\frac{ik}{2z}(x^2 + y^2)\right) \\ \times \mathcal{F}\left[U(x_0, y_0, 0) \exp\left(\frac{ik}{2z}(x_0^2 + y_0^2)\right)\right] \quad (3)$$

where $(x, y, z)$ denotes the observation plane, $k = 2\pi/\lambda$, $\mathcal{F}[\ ]$ is the Fourier transform. Under the condition of the coefficient $m = 1$, $n = 0$ and the $\omega_0 = 0.2$mm, the generation of high-order OV via HG beams with the CP is simulated in Fig. 1 by the substitution of Eq.(2) into Eq.(3). The intensity distributions dramatically change from HG beams to LG beams at different distances (z = 0m, 1m, 2m, 3m), shown in Fig. 1(a). Accordingly, the phase patterns gradually change to a phase distribution of LG beams as depicted in Fig. 1(b). It should be noted that, the phase pattern in the first column is a phase pattern of HG beams with a CP. Since the light field has not propagated (z = 0m) and the CP hasn't affected the light field yet, the corresponding intensity distribution remains a kind of HG distribution.

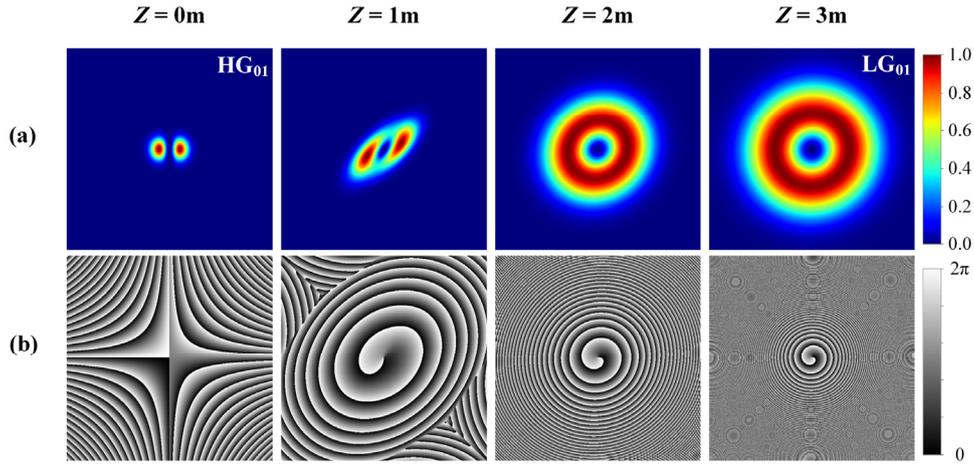

Fig. 1. The generation of high-order OV via HG beams with the CP of waist radius 0.2mm at different distances (z=0m, 1m, 2m, 3m). (a) simulated intensity distributions, (b) corresponding phase patterns.

## B. Measurement of high-order OV

Akin to the generation of high-order OV in the situation where the HG beams carry the CP (cf. Fig. 1), now we consider adopting the CP to implement the measurement of high-order OV. Without loss of generality, we take following input as an example

$$U_M(r_0, \phi_0, 0) = \sqrt{\frac{2p!}{\pi(p+|l|)\omega_0^2}} \left(\frac{\sqrt{2}r_0}{\omega_0}\right)^{|l|} L_p^{|l|}\left(\frac{2r_0^2}{\omega_0^2}\right) \exp\left(-\frac{r_0^2}{\omega_0^2}\right) \exp(-il\phi_0) \times \exp(i\psi) \quad (4)$$

where $L_p^{|l|}$ denotes the Laguerre polynomial. The relationship between parameters of HG and LG beam is

$$l = |n - m|$$
$$p = \min(n, m) \tag{5}$$

where $l$ is the TC and $p$ denotes the RN of the generated HG beams.

By substituting Eqs.(4) into Eqs.(3), the measurement of high-order OV via LG beams with the CP is simulated in Fig. 2, under the condition of the coefficient $u = 30$, $l = 1$, $p = 0$ and the $\omega_0 = 0.2$mm. The evolution of intensity distributions from LG beams (donut distribution) to HG beams (two lobes distribution) at different distances (z = 0m, 1m, 2m, 3m) is shown in Fig. 2(a). The corresponding phase patterns gradually change to a phase distribution of HG beams as depicted in Fig. 2(b). It is conceivable that, as the propagation distance increases, the change of the intensity and phase will be a reverse process of Fig. 1.

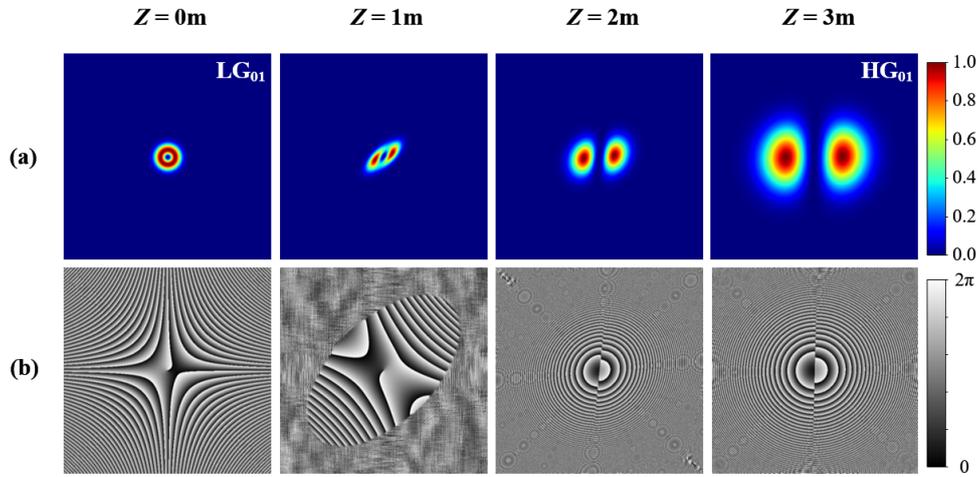

Fig. 2. The measurement of high-order OV via LG beams with the CP of waist radius 0.2mm at different distances (z=0m, 1m, 2m, 3m). (a) simulated intensity distributions, (b) corresponding phase patterns.

In addition, by reason of the normalization processing, the intensity distributions of Fig. 1(a) and Fig. 2(a) do not seems to weaken as increasing the propagation distance (the same hereinafter).

## 3. Experimental results

The experimental setup for generation and measurement of high-order OV by utilizing the CP is shown in Fig. 3. The laser delivers a collimated Gaussian beam with wavelengths of 632.8nm after a half-wave plate (HWP), a linear polarizer (LP) and a telescope consists of two lenses (L1, L2) are used for collimation. The combination of the LP and the HWP is served to rotate the laser polarization state along the long display axis of SLM (by setting the LP polarization in the vertical direction) and adjust the power of an incident light on SLM (by rotating the HWP). The SLM (HOLOEYE PLUTO-NIR-011) precisely modulate the incident light via loading a hologram and then the aperture (AP) is used to select the first diffraction order of the beam to avoid other stray light. The beam is adjusted to an appropriate size through another telescope (L3, L4) and the intensity pattern is registered by a CCD camera (NEWPORT LBP2).

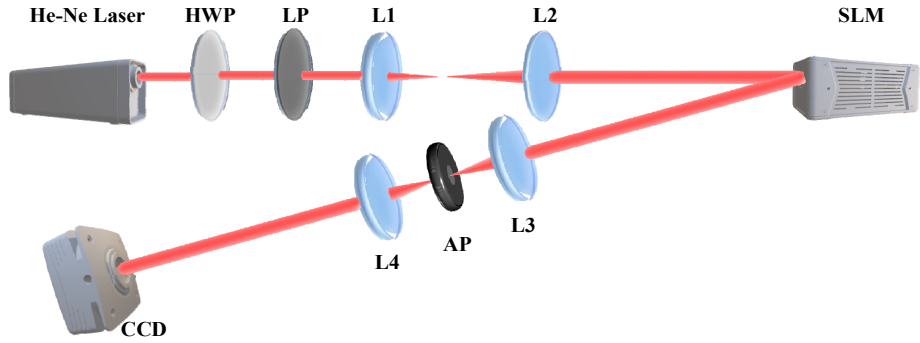

Fig. 3. Experimental setup for generation and measurement of high-order OV by utilizing the CP. HWP: half-wave plate. LP: linear polarizer. L: lens. SLM: spatial light modulator. AP: aperture. CCD: charge couple device.

To ensure the experimental accuracy and improve the quality of light field, it is necessary to calculate a hologram could precisely modulate the light field. Taking the measurement of LG beams as an example, the calculation process of a hologram is depicted in Fig. 4. First, we calculate the hologram of a LG beam with $l = 10$, $p = 5$ as shown in Fig. 4(a) via employing the complex amplitude modulation, which can be utilized to modulate the phase and intensity of the incident light simultaneously with only a pure phase SLM[16]. Next, the hologram of the LG beam is superimposed with a CP (shown in Fig. 4(b) to measure the LG beams) and a blazed grating (shown in Fig. 4(c) to select the first diffraction order), thereby obtaining a hologram that can accurately modulate the light field as Fig. 4(d) shows.

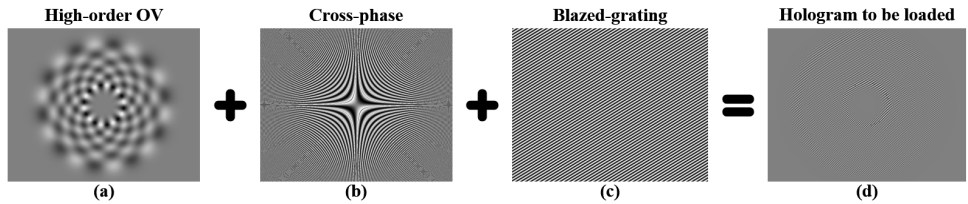

Fig. 4. Schematic representation of the hologram calculation. (a) the hologram could modulate the phase and intensity of incident light via employing complex the amplitude modulation. (b) the CP distribution. (c) the phase distribution of a blazed grating. (d) the ready to load hologram.

### A. Experimental generation of high-order OV

Based on the setup shown in Fig. 3, we first consider the generation of high-order OV in the situation where the HG beams carry the CP with the coefficient $u = 50$. The simulated intensity distributions of HG beams adopted in the experiment are depicted in Fig. 5(a), the corresponding parameters are $n = 40$, $m = 5$; $n = 50$, $m = 10$; $n = 80$, $m = 20$, respectively. The generated LG beams via utilizing the HG beams (shown in Fig. 5(a)) with the CP are simulated in Fig. 5(b). After loading the specific hologram which contains the CP, the blazed grating and the phase of HG beams, we have achieved the conversion from HG beams to LG beams, which means realizing the generation of high-order OV, as shown in Fig. 5(c). We obtain the parameters of LG beams which are $l = 35$, $p = 5$; $l = 40$, $p = 10$; $l = 60$, $p = 20$ in turn via Eqs. (5). As the order of generated LG beams increases, the power and the size of the main bright ring decreases gradually. The experimental results are basically consistent with the simulation outcomes. Nevertheless, it is to be noted that the resulting LG beams are not ideal and their intensity distributions are slightly elliptical, and more details will be further discussed below.

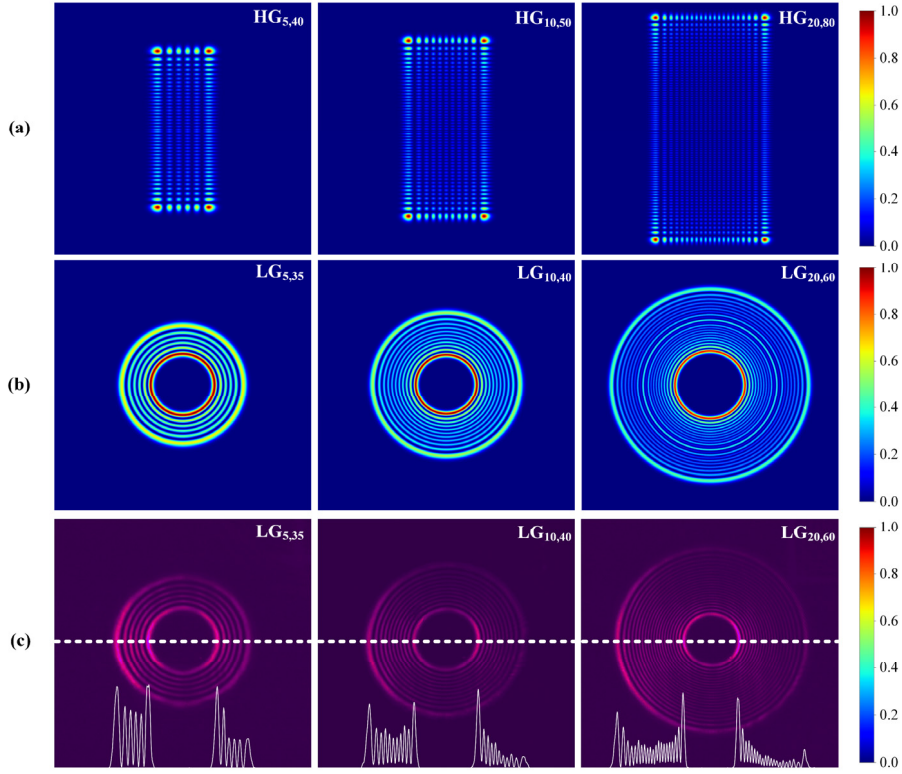

Fig. 5. Generation of high-order OV via employing the CP in the far-field. (a) Simulated intensity distributions of HG beams which are served to implement the generation of high-order OV, the corresponding parameters are $n=40$, $m=5$; $n=50$, $m=10$; $n=80$, $m=20$, respectively. (b) Simulated intensity distributions of the generated LG beams via utilizing the corresponding HG beams with the CP. (c) Experimental intensity distributions of the generated LG beams, and the parameters are obtained that $l=35$, $p=5$; $l=40$, $p=10$; $l=60$, $p=20$ in turn via Eqs. (5).

## B. Experimental measurement of high-order OV

Then we consider the measurement case of high-order OV in the situation where the LG beams carry the CP with the coefficient $u=50$. The LG beams used in the experiment are simulated in Fig. 5(a), and the generated HG beams via utilizing the LG beams (shown in Fig. 6(a)) with the CP are simulated in Fig. 6(b). With the same method, the CP is employed to achieve the conversion from LG to HG beams, achieving the measurement of high-order OV, as shown in Fig. 6(c). We calculate the parameters of HG beams which are $n=101$, $m=1$; $n=110$, $m=10$; $n=200$, $m=0$ in turn. According to the Eqs.(5), the parameters of LG beams before conversion can be derived as $l=100$, $p=1$; $l=100$, $p=10$; $l=200$, $p=0$, respectively.

The experimental results are basically consistent with the simulation outcomes. However, whether it is a simulation or an experimental result, the converted HG beams have a certain degree of tilt, which we will discuss below.

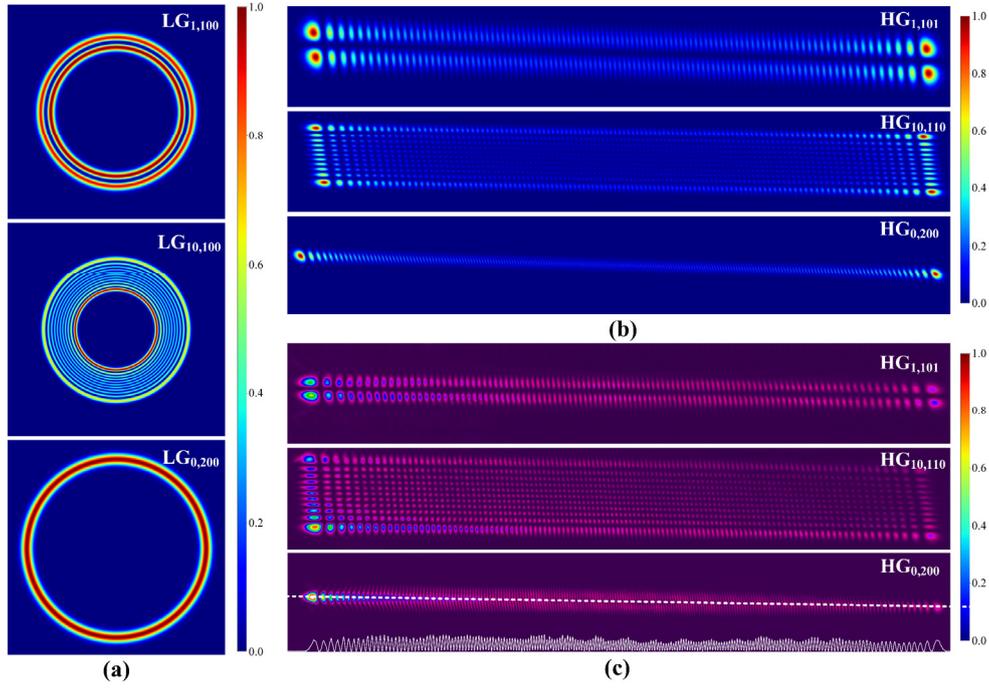

Fig. 6. Measurement of high-order OV via employing the CP in the far-field. (a) Simulated intensity distributions of LG beams which are served to implement the measurement of high-order OV. (b) Simulated intensity distributions of the generated HG beams via utilizing the corresponding LG beams with the CP. (c) Experimental intensity distributions of the generated HG beams, and the parameters are obtained that $n=101$, $m=1$; $n=110$, $m=10$; $n=200$, $m=0$. The parameters of LG beams before conversion can be derived as $l=100$, $p=1$; $l=100$, $p=10$; $l=200$, $p=0$ in turn via Eqs. (5).

In addition to measuring the orders of higher-order OV, the CP can also be served to measure the sign of TCs. For simplicity, we discuss the case of measuring the OVs with $l=10$ and $l=-10$, as depicted in Fig. 7. With the increase of the propagation distance, two OVs with opposite TC signs gradually approximate an orthogonal distribution at the same distance. Herein, the CP is regarded as a cylindrical lens. It is conceivable that the OV of $l=10$ and $l=-10$ with CP can yield $HG_{0,10}$ and $HG_{10,0}$ via $\pi/2$ phase difference introduced by Gouy phase[17]. That's why the CP can be employed to measure both value and sign of TC.

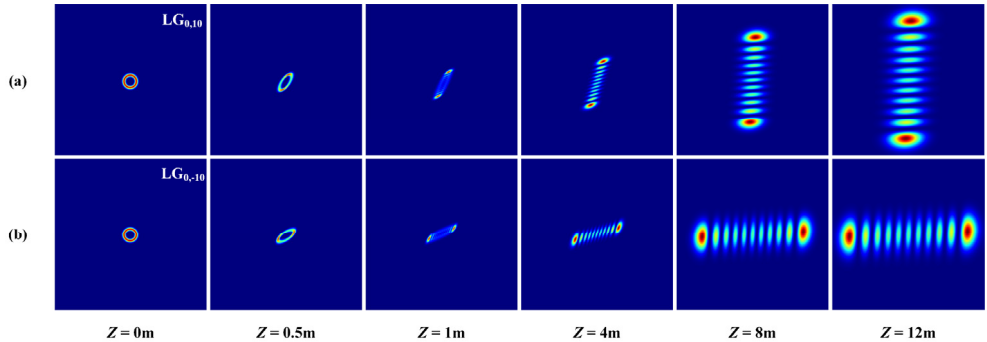

Fig. 7 Sign measurement of OV's TC via the CP where the waist radius of initial OV is 0.5mm and the TCs are (a) $l=-10$ and (b) $l=10$.

In fact, in the process of utilizing the CP to implement the generation and measurement of high-order OV, the coefficient $u$ needs to be adjusted to meet the experimental requirements for different initial conditions, which is similar to the cylindrical lens has the following requirement for the waist radius $\omega_0$ of the incident light[18]

$$\omega_o = \sqrt{\frac{\left(1+\sqrt{2}\right)f\lambda}{\sqrt{2}\pi}} \tag{6}$$

where $k = 2\pi/\lambda$, $f$ denotes the focal length and $\lambda$ denotes the wave length of the incident light. Thus, if $k/(2f)$ is set to $u$, we can derive the expression of $u$ as

$$u = \frac{1+1/\sqrt{2}}{\omega_o^2} \tag{7}$$

However, actually, one CP cannot be simply equivalent to a cylindrical lens. The phase term of a cylindrical lens can be expressed as Eqs.(8) when settled vertical.

$$\psi_p^{90°} = \exp\left(ix^2 \frac{k}{2f}\right) \tag{8}$$

which can be expanded into

$$\begin{aligned}\psi_p^{90°} &= \exp\left[\frac{ik}{2f} \times \left(\frac{1}{2}(x^2+y^2) + \frac{1}{2}(x^2-y^2)\right)\right] \\ &= \exp\left\{\frac{ik}{2f} \times \left[\frac{1}{2}(x^2+y^2) + \left(\frac{\sqrt{2}}{2}x - \frac{\sqrt{2}}{2}y\right)\left(\frac{\sqrt{2}}{2}x + \frac{\sqrt{2}}{2}y\right)\right]\right\}\end{aligned} \tag{9}$$

The term $\frac{1}{2}(x^2+y^2)$ can be considered as a spherical lens, and if we take Eqs.(1) into consideration, we can get

$$\psi_p^{90°} = \exp\left\{\frac{ik}{2f} \times \left[\frac{1}{2}(x^2+y^2) + \left(x\cos\frac{\pi}{4} - y\sin\frac{\pi}{4}\right)\left(x\sin\frac{\pi}{4} + y\cos\frac{\pi}{4}\right)\right]\right\} \tag{10}$$

Namely, a vertically placed cylindrical lens can be equivalent to one spherical lens plus one CP which is rotated by 45 degrees, which is the reason why one CP cannot be simply equivalent to a cylindrical lens.

### 4. Discussion

**A. Waist radius and generation of high-order OV**

We noticed that the waist radius of initial HG beams does affect the generation of high-order OV. An interesting finding is that the larger the radii, the larger the ellipticity. Fig. 8 shows the intensity distributions of LG beams corresponding to the different waist radii of initial HG beams after 4m transmission. In order to more intuitively reveal the relationship between the radius and ellipticity, we calculate the ellipticity $\Gamma$, which is according to

$$\Gamma = \frac{\sqrt{\max(D)^2 - \min(D)^2}}{\max(D)} \tag{11}$$

where $D$ is all the diameter of a certain LG beam, the ellipticity $\Gamma$ of generated LG beams versus the waist radii of initial HG beams is calculated, as shown in Fig. 10(a). Wherein, the dashed line denotes the simulation results and the real line denotes the experimental findings. The results indicate that the ellipticity of LG beams increases with the increase of the radius, despite the increase range gradually decreases.

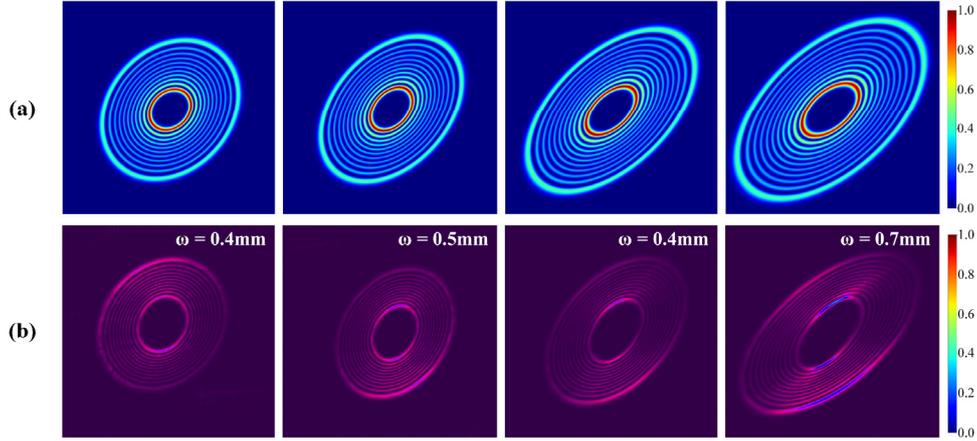

Fig. 8. Generated LG intensity distributions with different waist radii of initial HG beams at 4m, which reflects the effect of waist radius on generation of high-order OV. Waist radii of initial HG beams from the first column to the fourth column is (a)0.4mm, (b)0.5mm, (c)0.6mm, (d)0.7mm, where (a) are simulation results and (b) are experimental results.

## B. Waist radius and measurement of high-order OV

Similarly, the choice of the waist radius will also have an effect on the measurement of high-order OV. The larger the waist radius of LG beams, the greater the slope of the converted HG beams. Fig. 9 shows the intensity distributions of HG beams at the different waist radii of initial LG beams after 4m transmission. In order to more intuitively reflect the relationship between radius and slope, we calculate the slope K as shown in Fig. 10(b). The results demonstrate that the slope is positively correlated with the waist radius of initial LG beams, and the experimental results are basically consistent with the simulation results.

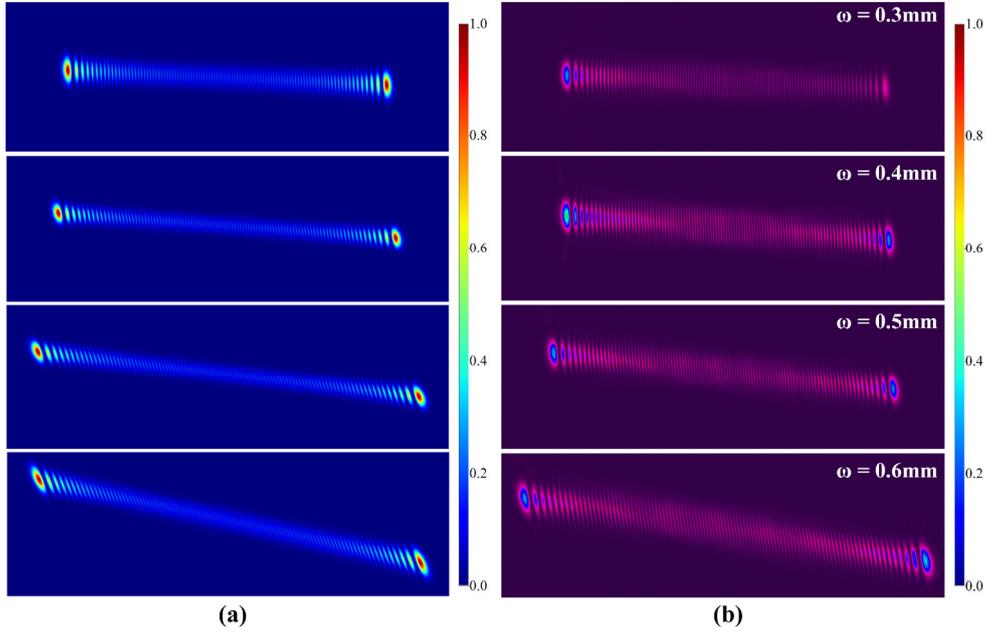

Fig. 9. Intensity distributions of HG beams with different waist radii of initial LG beams at 4m, which reflects the effect of waist radius on measurement of high-order OV. Waist radii of initial LG beams from the first row to the fourth row is (a)0.3mm, (b)0.4mm, (c)0.5mm, (d)0.6mm, where (a) are simulation results and (b) are experimental results.

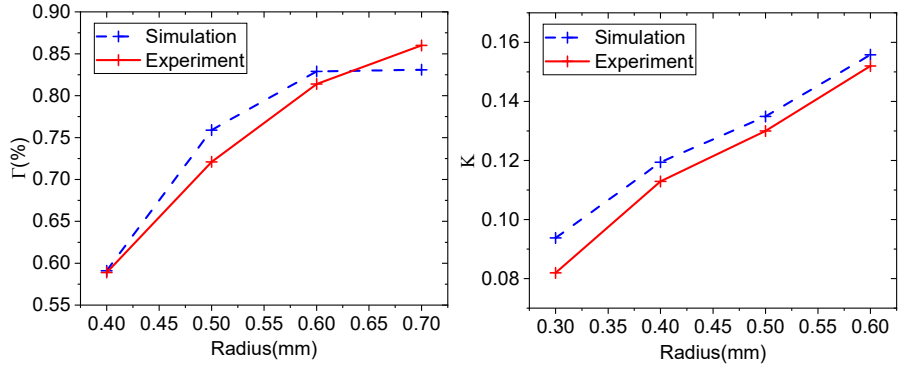

Fig. 10. Ellipticity and slope curves at the different waist radii: (a) ellipticity $\Gamma$ versus waist radius of initial HG beams, (b) slope $K$ versus waist radius of initial LG beams.

The reason for the above two cases happened is mainly because the CP is not evenly distributed, therefore, the degree of modulation of the same beam in different place is not the same. The phase change is small near the origin, while is large away from the origin, as shown in Fig. 4(a). Consequently, the change caused by the CP to a large-radius beam is greater than a small one. The two phenomena related to waist radii mentioned above can also be quantitatively explained(cf.Eqs.(7)). We know that $u$ controls the conversion rate that uniquely determines the light field distributions in a certain plane (which is specifically reflected in the slope or ellipse of a beam, etc.), where the waist radius uniquely determines the $u$ as well. As the waist radius increases, $u$ gradually decreases, while the larger the $u$, the faster the conversion rate. Therefore, the larger the waist radius of initial beams, the lower the conversion rate of the beams, which is specifically reflected in the same plane: the beam with an initial small waist radius is much closer to the completion of the conversion (with a smaller slope or ellipse). For a beam with a large waist radius, a longer distance is required to complete the same conversion. In other words, the reason why we adjust $u$ is actually to match the CP with the waist radius of the incident light.

### C. Confirmatory experiment

We try to change the parameters of HG and LG beams in the generation and measurement of high-order OV to verify our explanations above. It shows that changing the HG mode number does not affect the ellipticity of resulting LG beams while keeping the waist radius constant. As shown in Fig. 11(a), under the condition of an initial radius of 0.3 mm, the intensity distributions of the LG beams converted from HG beams with mode numbers $m = 5$, $n = 20$, 40, 60, and 80 are shown from top left to bottom right, respectively. Although the number of modes is different, the ellipticity remains unchanged at 0.56. Similarly, changing $l$ or $p$ of the LG beams does not affect the slope of the resulting HG beams. Fig. 11 (b) from left to right are the intensity distributions of the HG beams converted from LG beams with $p = 5$, $l = 20$, 40, 60, 80, respectively, under the condition that the waist radius is 0.3mm. Despite the TCs are different, the slope is stable at 0.22.

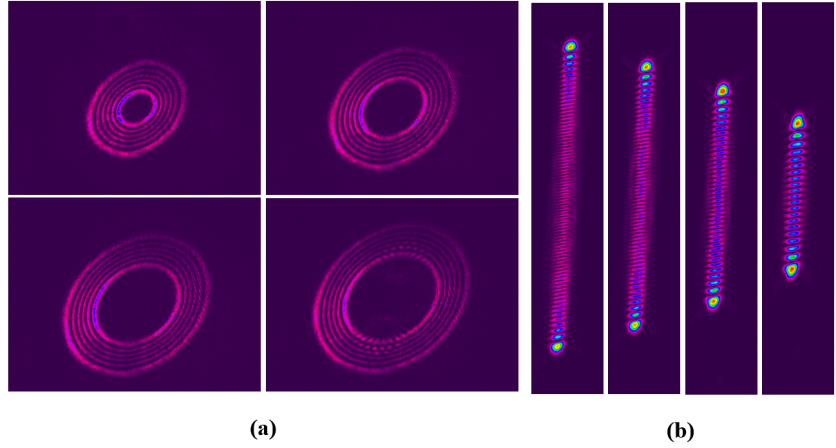

(a)    (b)

Fig. 11. Experimental results of intensity distributions of LG and HG with the waist radius of 0.3mm at 4m. (a) initial parameters of HG beams: $m = 5$, $n = 20, 40, 60,$ and $80$; (b) initial parameters of LG beams: $p = 5$, $l = 20, 40, 60, 80$.

## 5. Conclusion

In conclusion, a generation and measurement method of high-order OV via utilizing the CP is investigated and the schematic of generating a hologram by the complex amplitude modulation is introduced. In the experiment, we generate OVs with $l = 35$, $p = 10$; $l = 40$, $p = 10$; $l = 60$, $p = 20$ and successfully measure OVs with $l = 100$, $p = 1$; $l = 100$, $p = 10$; $l = 200$, $p = 0$, respectively. Regardless of generation and measurement, experimental results agree well with simulation outcome. On this basis, the distributions of LG and HG beams (corresponding to the generation and measurement) versus waist radius of initial light beam is discussed. It is found that waist radius does have an impact on the intensity distributions: the larger the waist radius of initial beams, the greater the slope of the converted HG beams and the larger the ellipticity of the converted LG beams. However, once the radius is settled, the shape (the ellipticity or slope) of the intensity distributions remains unchanged even though other parameters are altered.

This work provides an alternative method to generate or measure high-order OV, which will facilitate applications in optical micro-manipulation, quantum entanglement and rotation speed detection.